# Network-based Biased Tree Ensembles (NetBiTE) for Drug Sensitivity Prediction and Drug Sensitivity Biomarker Identification in Cancer

Ali Oskooei[1], Matteo Manica[1,2], Roland Mathis[1] and María Rodríguez Martínez[1]

[1] IBM Research – Zurich, Säumerstrasse 4, 8803 Rüschlikon, Switzerland

[2] Institute für Molekulare Systembiologie, Auguste-Piccard-Hof 1 8093, Zürich, Switzerland

**Abstract**

We present the Network-based Biased Tree Ensembles (NetBiTE) method for drug sensitivity prediction and drug sensitivity biomarker identification in cancer using a combination of prior knowledge and gene expression data. Our devised method consists of a biased tree ensemble that is built according to a probabilistic bias weight distribution. The bias weight distribution is obtained from the assignment of high weights to the drug targets and propagating the assigned weights over a protein-protein interaction network such as STRING. The propagation of weights, defines neighborhoods of influence around the drug targets and as such simulates the spread of perturbations within the cell, following drug administration. Using a synthetic dataset, we showcase how application of biased tree ensembles (BiTE) results in significant accuracy gains at a much lower computational cost compared to the unbiased random forests (RF) algorithm. We then apply NetBiTE to the Genomics of Drug Sensitivity in Cancer (GDSC) dataset and demonstrate that NetBiTE outperforms RF in predicting IC50 drug sensitivity, only for drugs that target membrane receptor pathways (MRPs): RTK, EGFR and IGFR signaling pathways. We propose based on the NetBiTE results, that for drugs that inhibit MRPs, the expression of target genes prior to drug administration is a biomarker for IC50 drug sensitivity following drug administration. We further verify and reinforce this proposition through control studies on, PI3K/MTOR signaling pathway inhibitors, a drug category that does not target MRPs, and through assignment of dummy targets to MRP inhibiting drugs and investigating the variation in NetBiTE accuracy.

## 1. Introduction

There is strong evidence that the tumor's genetic makeup can influence the outcome of anti-cancer drug treatments (1, 2), resulting in heterogeneity in patient clinical response to therapeutic drugs (3). This varied clinical response has led to the promise of personalized (or precision) medicine in cancer, where molecular biomarkers, e.g. gene expression, obtained from a patient's tumor profiling may be used to design a personalized course of treatment. Targeted treatments have been shown to improve survival rates, for instance, in treating chronic myeloid leukemia (BCR–ABL) and malignant melanoma (BRAF) (4,5). Despite these success stories, variability in drug response still remains an open challenge and the link between genetic and epigenetic alterations and drug response is not appropriately characterized for a large number of cancer drugs (6,7). As large datasets emerge containing genetic profiles of tumors and their associated drug sensitivity, there is a need for computational methods that can effectively harness the available data and link genetic profiles with drug sensitivity through identification of important biomarkers (8–13).

The Sanger Institute's Genomics of Drug Sensitivity in Cancer (GDSC) database is a vast resource of over 200 cancer compounds screened with over a thousand genetically profiled pan-cancer cell lines (6). The dataset has been of particular interest for drug sensitivity prediction and biomarker identification efforts (3,8,14–18). These include a number of works employing quantitative, statistical and machine learning methods such as : Cell line-similarity and drug-similarity based models (19) multilevel mixed effect models using all drug-cell line combinations (20), quantitative structure-activity relationship (QSAR) analysis using kernelized Bayesian matrix factorization (21), lasso and elastic net models for drug sensitivity prediction and target identification (8,22,23), as well as logic models for predictor identification (24).

In this work, we introduce a novel machine learning method that enables us to predict IC50 values and identify informative predictors for drug sensitivity using the GDSC dataset. Our approach is based on constructing a biased tree ensemble, where bias is elaborately designed to recapitulate the prior knowledge of drug targets and their high-confidence biomolecular interactions extracted from the STRING database of molecular interactions (25). Tree ensemble methods (26,27) such as the popular random forests (RF) (28) algorithm consist of an aggregation of decision trees and are suitable for dealing with high-dimensionality (29) (i.e. small number of samples and large number of features) that is often encountered in biomolecular datasets. In addition, unlike regularized linear methods such as lasso and elastic net (30), regression trees can capture non-linear relationships. Furthermore, tree ensemble methods are robust and have few tuning parameters (number of trees, *mtry*, and tree depth) and as such are easy to train. Due to these favorable attributes, tree ensembles, and in particular the random forests algorithm, have been used extensively for the analysis of biomolecular data (31–36).

In this paper, we first introduce the Biased Tree Ensembles (BiTE) approach, where the classification and regression trees (CART) (37) are constructed according to prior knowledge. Unlike random tree ensembles (i.e., RF) in which all features have an equal probability of being selected as split variables in a tree (figure 1A), in BiTE, features that are more important or informative according to the available prior knowledge are given a higher probability (figure 1B). We demonstrate that BiTE is a more transparent and interpretable algorithm compared to RF, as it is immediately clear which set of features contributed the most to the model performance. For instance, if a set of features results in BiTE's loss of accuracy, it can be deduced that the features were uninformative predictors; conversely, an improved accuracy can be attributed to the set of features towards which we biased the model. In this manner, BiTE may be used to examine the predictive power of various features in a transparent and controllable manner.

Building upon BiTE, we propose the Network-based Biased Tree Ensembles (NetBiTE) algorithm, where two layers of prior knowledge – instead of one in BiTE – are fed into the model. First, drug target proteins are determined from drug databases and the literature and are assigned an initial bias weight. Second, the initial bias weights are propagated over STRING, a network of protein-protein-interaction (PPI) comprising the entire gene set. A number of network-based methods have been previously put forward that take advantage of PPI networks in combination with biomolecular profiles of cells, in order to identify subnetworks that represent a pathway or a functional complex (38,39). Network propagation (or diffusion) over PPI networks has been previously used to identify pathways, subnetworks or associations that represent a disease, a tumor type or a patient (40–42). Network propagation in essence defines a "neighborhood of influence" surrounding an entity of interest, for instance a mutated gene (42). NetBiTE utilizes network diffusion over STRING PPI network in order to establish a neighborhood of influence surrounding the drug target proteins and construct tree ensembles that are biased towards this neighborhood. Even though several modified random forests or tree ensembles algorithms have been previously proposed (43–46), to the best of our knowledge, NetBiTE is the first algorithm in which multiple layers of prior knowledge are quantitatively and systematically combined and utilized in constructing biased tree ensembles.

In the following sections, we demonstrate that BiTE and NetBiTE outperform RF in predicting IC50 drug sensitivity using both a synthetic dataset and the GDSC dataset. In addition, we showcase how NetBiTE in conjunction with the GDSC dataset and the STRING PPI network can identify important biomarkers for drug sensitivity.

The organization of this paper is as follows. In section 2.1, we compare BiTE versus RF using a synthetic dataset. We showcase that BiTE can achieve a superior performance and stability at a significantly lower computational cost. In section 2.2, we apply NetBiTE to the GDSC data for a panel of 50 cancer drugs and compare the predictive performance with that of RF. We demonstrate that NetBiTE achieves significant accuracy gains over RF for drugs than inhibit membrane receptor pathways (MRPs), suggesting that the expression of their reported target genes is an informative biomarker for drug sensitivity. We further investigate this hypothesis by studying all drugs within the GDSC database that target MRPs and by performing two control experiments. In section 3, we discuss the possible reasons behind our observations in the context of prior findings related to the role of oncogenes in cancer development as well as drug sensitivity and resistance.

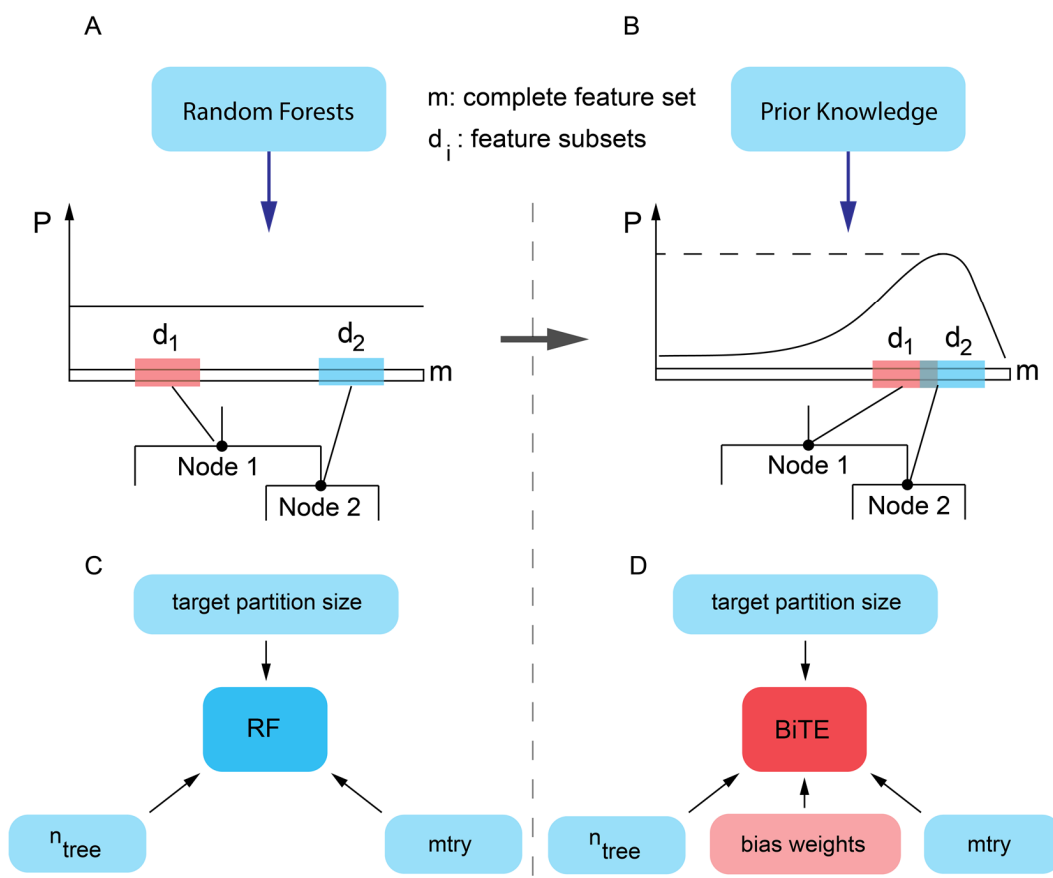

**Figure 1. Working principles of biased tree ensembles (BiTE) compared against random forests (RF).** A, B) diagrams describe the standard random forests (RF) algorithm and our devised biased tree ensembles (BiTE) approach. At each node of the tree, both algorithms draw a subset of features (di) from which they select a split feature through optimization of a loss function. In RF, the feature subset is selected at random while BiTE biases the selection towards more informative features according to prior knowledge. C, D) Tuning parameters for RF and BiTE algorithms. There are three tuning parameters for RF: the number of trees, $n_{tree}$, the target partition size (*TPS*) – the minimum number of samples in the leaf nodes of the tree, and the number of features to consider when looking for the best split, *mtry*. In BiTE, there is an additional tuning parameter, the bias weight distribution that controls the probability of each feature being included in the feature subset ($d_i$) at each split of the tree. If weights are assigned according to informative prior knowledge, BiTE results in significant performance gains at a lower computational cost.

## 2. Methods and Materials

### 2.1 Dataset

Throughout this work we made use of the gene expression and drug IC50 data publicly available as part of the Genomics of Drug Sensitivity in Cancer (GDSC) database. The Genomics of Drug Sensitivity in Cancer Project is part of a collaboration between The Cancer Genome Project at the Wellcome Trust Sanger Institute (UK) and the Center for Molecular Therapeutics, Massachusetts General Hospital Cancer Center (USA)(1,2).

The dataset includes screening results of more than a thousand genetically profiled human pan-cancer cell lines with a wide range of anti-cancer compounds (265 compounds as of the writing of this paper). The screened compounds include chemotherapeutic drugs as well as targeted therapeutics from various sources (2). We based our models on gene expression data as it has been shown to be more predictive of drug sensitivity in comparison to genomics (i.e., copy number variation and mutations) or epigenomics (i.e., methylation) data (47).

The synthetic dataset used in the case study of figure 2 was created by randomly selecting a subset of 200 cell lines and 200 genes from the standardized RMA basal expression profiles within the GDSC database. Synthetic IC50 values were generated from two randomly selected "drug target" genes according to the following 5$^{th}$ degree relationship:

$$IC50_{synthetic} = aX_i^5 + bX_j^5 + cX_i^3 X_j^2 + dX_i^2 X_j^3 \qquad i,j \in \{\mathbb{N} < 200\} \qquad (1)$$

A panel of 50 cancer compounds that had been screened with the highest number of cell lines (highest number of samples, n=883) was selected for the initial characterization (supplementary figure S1) and RF versus NetBiTE comparison studies (figure 3). The drugs in the studied panel belonged to twelve different categories of drugs (figure 3C).

Studies on individual drug categories were performed for four different drug categories targeting four different signaling pathways. The four drug categories, the number of cell lines screened ($n$) within each category, and the number of genes within the dataset ($m$) are: RTK signaling pathway inhibitors (RSPIs) drugs (21 drugs, $n$ = 310, $m$ = 15698), EGFR signaling pathway inhibitors (ESPIs) drugs (7 drugs, $n$ = 309, $m$ = 15698), PI3K/MTOR signaling pathway inhibitors (PMSPIs) drugs (21 drugs, $n$ = 238, $m$ = 15698) and IGFR signaling pathway inhibitors (ISPI) drugs (4 drugs, $n$ = 820, $m$ = 15698). All RMA basal expression profiles were standardized and IC50 drug response data were scaled between zero and one.

**2.2 Methods**

We have developed a new methodology for drug sensitivity prediction using a combination of gene expression data and prior knowledge. Our Biased Tree Ensemble (BiTE) algorithm consists of an ensemble of CART trees that are constructed according to a bias distribution that represents prior knowledge about the drug targets and their interactions and associations with other genes within the dataset. In the popular random forests algorithm (RF), a subset (d<m) of the total number of features is randomly selected and used to determine the next split variable that minimizes the variance (for regression) or Gini impurity (for classification) (28). For datasets with a large number of features, such as gene expression data, the optimization process can become computationally demanding (30).

Unlike the RF algorithm in which all features have an equal chance of being selected, in our method, we assign a weight to each feature (i.e., gene), that controls the probability of the feature being selected as a split variable. The variable selection scheme in BiTE, biases the selection of split variables towards a specified subset of variables (e.g. drug targets) that are known to be relevant or informative from prior knowledge. In addition to significantly reducing the computational cost, injection of informative prior knowledge renders the model more transparent and interpretable. Hence, BiTE can be used to verify the importance of a set of features by assigning a higher selection weight to the said features and investigating the variation in the model accuracy as compared to a model with uniform weights. Improvement or deterioration of the model performance, compared to the unbiased model, would indicate the predictive power of the studied feature set or lack thereof. Using this unique feature of BiTE, one can for instance, identify biomarkers for cancer drug sensitivity as we demonstrate in this work.

To construct the bias weight distribution to be ingested into the algorithm, we assign a high bias weight (e.g., $W$ = 0.6 or $W$ = 1, where $W$ = 0 means never select and $W$ = 1 means always select) to the reported drug target genes, while assigning a very small positive weight ($\varepsilon$ = 1e-5) to all other genes (figures 2B and 3A). For drug sensitivity prediction studies, we implement an additional step of propagating the initial weights distribution ($W_0$) over the STRING protein-protein-interaction (PPI) network. The STRING network contains protein-protein interaction information from various sources, including high-throughput lab experiments, computational predictions, automated text-mining of public text collections and information from additional databases (25). By propagating the initial weights over STRING, we smoothen the initial weights over a prior knowledge-based PPI network and thereby, introduce an additional layer of prior knowledge to our algorithm's weighting scheme. Our Network-based Biased Tree Ensembles (NetBiTE) algorithm is a result of combining BiTE with network propagation of bias weights.

The network smoothening of the weights can be described as a random walk and propagation of the initial drug target weights throughout the network (figure 3A). Let us denote the initial weights as $W_0$ and the string network as $S$ = ($P, E, A$), where $P$ are the protein vertices of the network, $E$ are the edges between the proteins and $A$ is the weighted adjacency matrix. The adjacency matrix weights indicate the level of confidence that a certain interaction exists. The smoothened weights are determined from an iterative solution of the propagation function (40,41,48):

$$W_{t+1} = \alpha W_t A' + (1-\alpha) W_0, \quad A' = D^{-\frac{1}{2}} A D^{-\frac{1}{2}}, \qquad (2)$$

where $A'$ is the normalized adjacency matrix – an adjacency matrix where the weight for each edge is normalized by the degrees of its end points – resulting in the probability of an edge existing between two nodes in a random walk over the network. $D$ is a diagonal matrix with the row sums of the adjacency matrix on the diagonal. The diffusion tuning parameter, $\alpha$ ($0 < \alpha < 1$), defines the distance that the prior knowledge weights can diffuse through the network. The higher $\alpha$ is, the more smoothened the resulting weights distribution will be. The optimal value of $\alpha$ depends on the network and is reported to be 0.7 for the STRING network (40). We further investigated the role of $\alpha$ and confirmed that for our drug sensitivity prediction studies $\alpha$ = 0.7 indeed results in the most desirable model performance (see supplementary section S2 and figure S2). Adopting a convergence rule of $e = (W_{t+1} - W_t) < $ 1e-6, we solved equation 2 iteratively for each drug and used the resultant bias weights distribution, $W_s$, in building a biased tree ensemble for IC50 prediction.

The propagation of the initial drug target weights over STRING simulates the propagation of perturbations within the cell following the drug administration. The biological linkage between the drug targets and their associated effectors is simulated via a random walk-based propagation of initial target weights over the PPI network, resulting in the spreading of the weights to the neighborhood of influence of a drug target, i.e. the set of high-confidence neighbors of the drug target.

In all IC50 prediction studies with NetBiTE and RF, we used a number of trees of $n_{tree}$ = 500 and a target partition size (*TPS*) of *TPS* = 1, as suggested by our tuning parameter analysis (see supplementary section S1). In addition, *mtry* was set to the number of reported targets for each drug. To evaluate the model performance, thirty to forty-fold cross validation (30 ≤ *k* ≤ 40) was used, based on the size of the dataset. Data samples were divided into *k* parts and at each step of the computation, one fraction of the data was left out as a test sample for model evaluation while the model was trained on the remaining data samples. This process was repeated until all subsets of the dataset were used once as test samples. The model performance was evaluated by determining the Pearson correlation, or the square root of the coefficient of determination ($\rho = \sqrt{R^2}$) between the actual IC50 values and the predicted ones (47,49). In IC50 prediction experiments, each complete round of computation was repeated 10 times and the mean of 10 trials was used as the representative model prediction. The repetition and averaging was used to minimize fluctuations inherent in all algorithms based on random sampling, e.g. RF and to lesser degree, NetBiTE (see figure 2 for comparison).

For computations performed in this work, we used Ranger random forests implementation in C++ (50) as well as the Scikit-learn random forests implementation (51). Ranger implementation accepts a split weights vector as an input and enables biased selection of split variables, necessary for NetBiTE computations. Data preprocessing, analysis and postprocessing were performed entirely in Python 3.5 and the plots and graphs were generated using the matplotlib Python library (52).

## 3. Results

### 3.1 Comparing BiTE with RF using a synthetic dataset

We performed a case study with a small synthetic dataset (200x200) to highlight the key differences and advantages of BiTE over RF in the context of drug sensitivity prediction. We created a synthetic dataset by extracting a subset of the GDSC rma normalized basal expression profiles; synthetic IC50 values were generated using a nonlinear drug sensitivity function based on two randomly selected targets (figure 2A).

Using the synthetic dataset, we compared the performances of three algorithms: i) standard random forests (RF); ii) BiTE with a weight of $W$ = 0.6 and a small positive weight ($\varepsilon$ = 1e-5) given to the two artificial drug targets and all other genes respectively – where $W$ = 0 means a feature is never selected and $W$ = 1 means it is always selected (see figure 2B); and iii) linear regression (LR) over the two target genes. The performances of all three methods are compared in figure 2C-2F.

In the case of RF and BiTE, we studied the model performance for various $n_{tree}$ ranging from 10 to 5000 trees. In addition, we varied *mtry* for each $n_{tree}$ from 2 (number of targets) to 200 (the total number of genes). As expected, for all investigated $n_{tree}$, RF achieves the highest performance when *mtry* reaches its maximum value of 200, as the model can test all available genes and pick the most informative. It is noteworthy that BiTE achieves the same maximal performance ($\rho \approx 0.9$) and stability with a minimal *mtry* (*mtry* = 2) even at tree numbers as low as ($n_{tree}$ = 10), which represents a significant saving in computational cost.

In all four subplots (figure 2C-2F), we have compared RF and BiTE with a linear regression model (the black dotted lines). As we have adopted a nonlinear drug sensitivity function, we do not expect LR to perform well, however it is insightful to include LR performance as a baseline. A conclusion that can be immediately drawn from the plots is that for very low *mtry* or $n_{tree}$, RF can underperform LR, even though the former is an inherently nonlinear model that should perform better in capturing a nonlinear relationship.

It is noteworthy that the results in figure 2 have been obtained using an ideal dataset with well-defined targets and IC50 function, resulting in an outstanding predictive performance using BiTE. Such significant improvement in model performance is not to be expected when using real IC50 data such as the GDSC data, where the informative predictors and the dependency of the IC50 values on these predictors is only partially known. However, if high quality prior knowledge is available, we do expect to see a clear improvement in model performance, stability and computational running times when compared to RF.

In the coming sections, we will further investigate whether our findings with the synthetic data can be generalized to real drug sensitivity data.

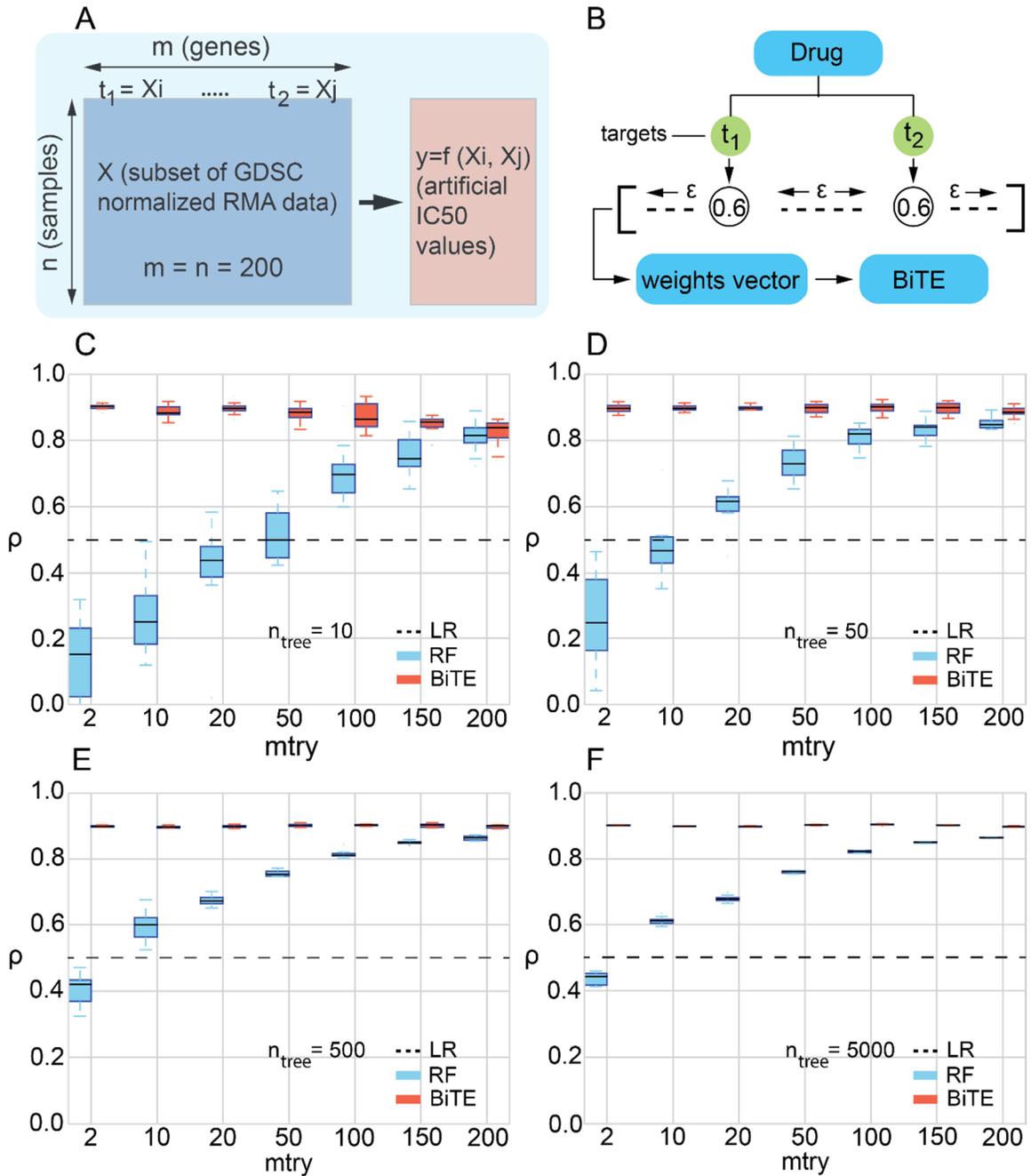

**Figure 2. BiTE, RF, and LR performances in predicting synthetic IC50 values in a synthetic dataset.** A) The synthetic dataset was created by selecting a random subset of normalized RMA data from the GDSC dataset. Synthetic IC50 values were generated using an arbitrary nonlinear function of two randomly selected genes adopted as synthetic drug targets. B) To demonstrate how BiTE exploits prior knowledge to outperform RF, we assigned a high bias weight

($w$ = 0.6) to synthetic targets while giving a very small positive weight, $\varepsilon$ = 1e-5, to all other features. As shown in (C)-(F), for $n_{tree}$ ranging from 10 to 5000, BiTE significantly outperforms RF at almost all tested *mtry* values. BiTE and RF performances converge only at very large *mtry*, when the splitting subset includes the entire feature set. In addition, BiTE performs consistently well at low *mtry* (as low as 2), drastically reducing the performance variability and computational running times. The dashed lines in the plots (marked with LR) represent the performance of a linear regression model trained on the synthetic targets. $\rho$ is the Pearson correlation between the predicted and actual IC50 values.

**3.2 Drug sensitivity prediction using network-based NetBiTE**

We applied our network-based Biased Tree Ensembles (NetBiTE) to the GDSC data for a panel of 50 cancer drugs. Unlike BiTE (section 2.1), where the bias weights are assigned solely based on our knowledge of drug targets (the first layer of prior knowledge), NetBiTE propagates the initial weight distribution over a network of molecular interactions (second layer of prior knowledge). Weight propagation results in a smoothened weight distribution where neighbors of known drug targets are also assigned a high weight (figure 3A).

As shown in figure 3A, for each drug in the studied panel of 50 drugs, a weight of one was given to each drug target and the resulting weight vector ($W_0$) was propagated over the STRING PPI network (25). As described in detail in section 2.2, the smoothening propagates the initial weights over the network resulting in lower weights for the targets ($W_{s,t}$ < 1) and positive weights for all other genes within the network ($W_{s,l}$ > 0). The smoothened weight distribution ($W_s$) is then fed to BiTE and the IC50 values for each drug are estimated.

In figure 3B, we study the effect of the tuning parameter, $\alpha$, that controls the diffusion depth of prior knowledge within the network, i.e. how much the known targets influence their neighbors. We compare NetBiTE's predictive accuracy with the optimal $\alpha$ = 0.7, as reported in the literature (40), and a much smaller value of $\alpha$ = 0.02. As shown in the plot, NetBiTE achieved a much higher accuracy with $\alpha$ = 0.7 ($\rho \approx$ 0.5) than with $\alpha$ = 0.02 ($\rho \approx$ 0.26). Hence, we adopted $\alpha$ = 0.7 in the remaining NetBiTE models.

Figure 3C shows the histogram for the drug categories within the panel of 50 studied drugs. Drugs targeting the PI3K/MTOR pathway, "Other" category, "Other Kinases" and drugs targeting the RTK signaling pathway are most represented. The varying representation in figure 3C is a result of the non-uniformity in the representation of various drug categories within the GDSC database.

The comparison between the predictive performance of NetBiTE and RF applied to the panel of 50 drugs is shown in figure 4D. At first glance the NetBiTE algorithm does not appear to be more accurate than RF. A closer inspection, however, points out a high variation in accuracy gain with NetBiTE across different drug categories. As shown in figure 3E, within the studied panel of 50 drugs, the majority of drugs that target RTK or EGFR signaling pathways show a noticeable improvement in prediction accuracy using NetBiTE while the other drug categories are either worsened or primarily unaffected by the use of NetBiTE. NetBiTE IC50 predictions for RTK and EGFR signaling pathway inhibitors experience the most frequent (60% and 50% of the drugs respectively) and significant improvements in accuracy with RTK inhibitors exhibiting a 25% improvement ($\Delta\rho$ = 0.12) and EGFR inhibitors, a 30% improvement ($\Delta\rho$ = 0.14). It is noteworthy that both RTK signaling and EGFR signaling pathways are membrane receptor pathways, suggesting a trend that we will investigate in the coming sections. The other drug categories experienced infrequent or insignificant improvements that do not reflect a statistically significant trend. Our preliminary observation is that drugs that target membrane receptor proteins achieve higher IC50 prediction accuracy with NetBiTE than with RF, suggesting that the expression of their target genes are informative biomarkers for IC50 drug sensitivity.

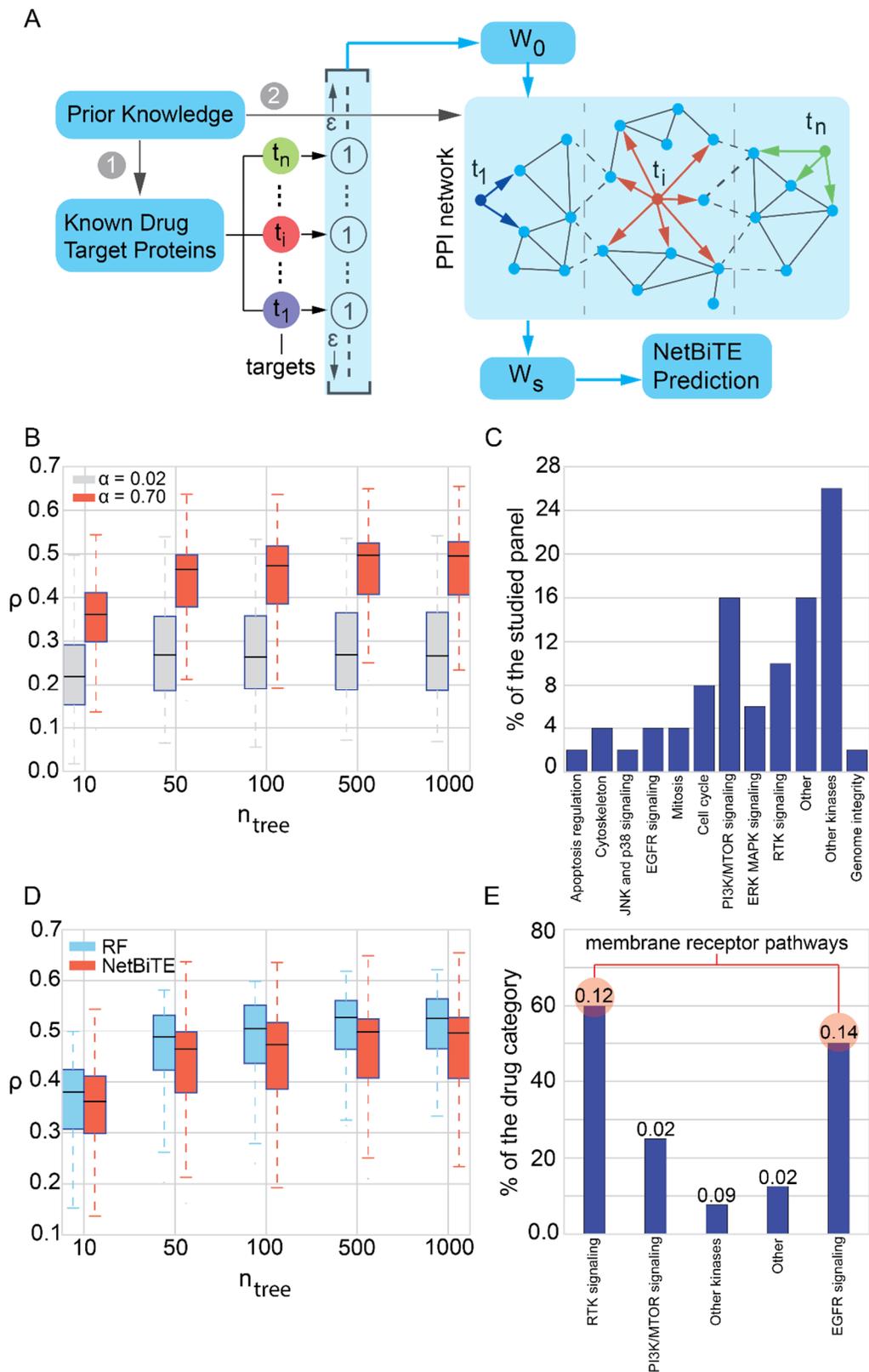

**Figure 3. IC50 drug sensitivity prediction with NetBiTE for a panel of 50 drugs identifies important drug target pathways.** A) The bias weights distribution utilized in NetBiTE was determined through network propagation of the

initial bias weights distribution, $W_0$. $W_0$ was obtained by assigning a weight of 1 to all drug targets and a small positive weight ($\varepsilon$ = 1e-5) to all other genes. Network propagation was performed via a random walk on the STRING PPI network. The propagated weights, $W_s$, were then fed to NetBiTE for IC50 prediction. B) Boxplots present the prediction accuracy achieved by NetBiTE using a $W_s$ obtained once with the optimal value of $\alpha$ = 0.7 and once with $\alpha$ = 0.02, accounting for a strong and insignificant weight propagation respectively. Network propagation with optimal $\alpha$ results in significant accuracy gains across the panel of 50 drugs ($\Delta\rho$ = 0.2 for $n_{tree}$ = 100) when compared with negligible propagation (p-value = 9.5e-14 given by two-sided t-test for $n_{tree}$ = 1000). C) Distribution of drug categories within the panel of 50 drugs. D) NetBiTE and RF performance comparison across the 50 drug panel. At first glance, it appears that NetBiTE is not offering any improvement over RF with pairwise t-test p-values across different trees ranging from 0.08 to 0.98. E) A closer look at the individual accuracy gain of various drug categories indicates that drugs that target RTK and EGFR signaling pathways (both membrane receptor pathways) experience significant accuracy gains at 25% ($\Delta\rho$ = 0.12) and 30% ($\Delta\rho$ = 0.14) improvement respectively. $\rho$ is the Pearson correlation between the predicted and actual IC50 values. The value on top of each bar in the bar plot represents the mean $\Delta\rho$ for that drug category.

To further investigate this hypothesis, we extracted all membrane receptor pathways inhibitors (MRPIs) within the GDSC database. The extracted drugs were inhibitors of three pathways: RTK signaling (21 drugs), EGFR signaling (7 drugs) and IGFR signaling (4 drugs) pathways. We then applied NetBiTE to the data for these three drug categories.

Figure 4A and 4B show the NetBiTE and RF results for RTK signaling pathway inhibitors (RSPIs). As demonstrated in figure 4A, NetBiTE is superior to RF particularly at lower tree numbers, as expected from the case study in 2.1. The difference between the prediction accuracies of NetBiTE and RF for 10 trees is statistically significant (p-value of 0.04 in a two-sided t-test). In figure 4B the difference in accuracies of NetBiTE and RF ($\Delta\rho = \rho_{NB} - \rho_{RF}$) is plotted for each RSPI drug. As shown, the majority of RSPI drugs fall under the positive region, meaning their IC50 prediction using NetBiTE was more accurate than RF predictions. 70% of RSPI drugs show improvement with NetBiTE for all numbers of trees with a maximum improvement of 0.3 for Linifanib. These results are in line with our earlier observations in figure 3E, where 60% of the RSPBI drugs, in a panel of 50, showed improvement in predictive performance using NetBiTE.

Figure 4C and 4D show the results for EGFR signaling pathway inhibitors (ESPIs). As shown in figure 4C, similarly to RSPI drugs, ESPI drugs show a noticeable improvement in prediction accuracy using NetBiTE.

Again, the improvement is more significant at lower tree numbers. Individual improvements in accuracy ($\Delta\rho = \rho_{NB} - \rho_{RF}$) are shown in figure 4D. Of the seven ESPI drugs, five show improvement with NetBiTE and four show consistent improvement across all $n_{tree}$.

The last category of MRPI drugs in the GDSC dataset are the drugs targeting the IGFR signaling pathway. Although there were only four drugs in this category, for the sake of thoroughness, we applied NetBiTE to these four drugs and compared the results with RF. The comparison between NetBiTE and RF performances for IGFR signaling pathway inhibitors (ISPIs) is shown in figure 4E. The results are similar to those shown before: accuracy improvement that is particularly significant at lower tree numbers. Individual gains in predictive performance ($\Delta\rho = \rho_{NB} - \rho_{RF}$) are shown in figure 4F. Three out of four drugs show a mean positive accuracy improvement across all tree numbers. However, due to the small number of drugs in this category we could not perform a meaningful statistical test to compare the two methods. The results however, do appear to be in line with the two other MRPI categories of drugs.

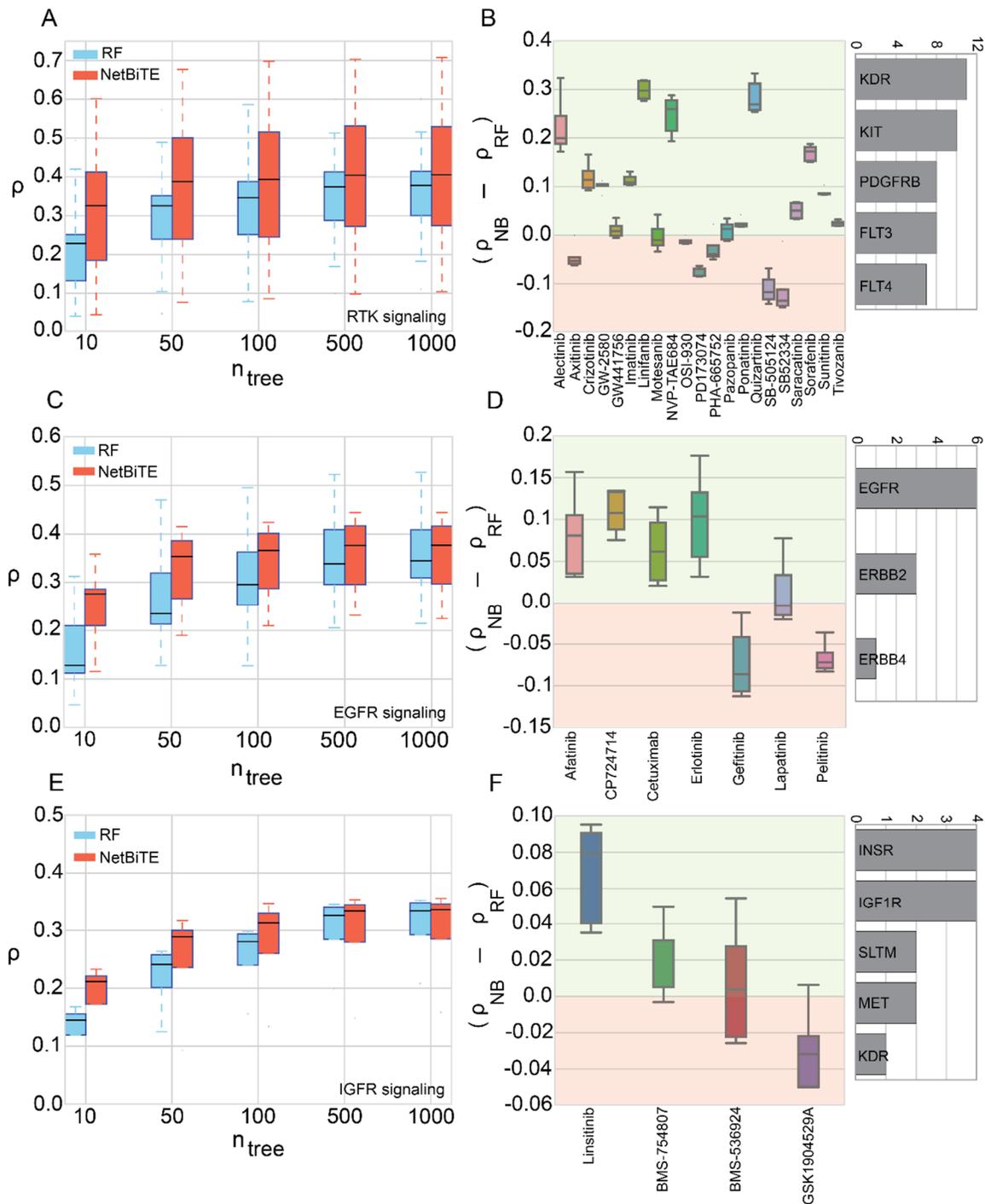

**Figure 4. NetBiTE versus RF performance comparisons for three categories of membrane receptor pathway inhibitors**, namely: RTK signaling, EGFR signaling and IGFR signaling pathway inhibitors. A) NetBiTE versus RF IC50 prediction results for the 21 RTK signaling pathway inhibitors within the GDSC dataset at various $n_{tree}$. The gain in accuracy using NetBiTE is particularly significant at low $n_{tree}$ (two-sided t-test p-value = 0.04 for $n_{tree}$ = 10). B) The individual gain in predictive performance for each drug at various $n_{tree}$ (five data points for each drug, one for each $n_{tree}$ studied in subplot A). As shown in (B), the model predictions for 75% of the RSPI drugs (15 out of 21) are improved using NetBiTE, with 14 out of the 15 drugs exhibiting a positive mean improvement across all $n_{tree}$. C) NetBiTE vs RF prediction performance across various $n_{tree}$ for all seven EGFR signaling pathway inhibitors within the GDSC dataset. The gain in performance is significant especially at lower tree numbers (two-sided t-test p-value of

0.02 for $n_{tree}$ = 10). D) Individual gain in performance for ESPI drugs. 71% of the drugs (5 out of 7) exhibit a gain in prediction accuracy with four drugs showing a positive mean improvement across all $n_{tree}$. E) NetBiTE versus RF IC50 prediction accuracy for ISPI drugs. Despite the low number of drugs in this category, a similar trend to RSPI and ESPI drugs can be observed. F) Individual improvements in accuracy for ISPI drugs. Three of the four drugs show a positive mean of improvement across all tree numbers. The histograms on the right show the top target genes according to the frequency of occurrence in each drug category.

By investigating the targets for the groups of drugs that show improved or worsened accuracy ($Δρ$ > 0 or $Δρ$ < 0) with NetBiTE, we observed that certain genes tend to appear more frequently in the targets of either the improved or worsened groups. As shown in the supplementary figure S3, in the case of RTK signaling pathway inhibitors (RSPIs), KIT, FLT3 and PDGFRB were the most frequent targets in improved accuracy group while KDR, TGFBR1 and FGFR1 were the most frequent targets in the worsened group of RSPI drugs. For ESPI drugs, ERBB2 was a target exclusively in the improved group and in the case of ISPI drugs, INSRR was a target exclusively in the drugs with improved accuracy.

The results in figures 3 and 4 consistently indicate that membrane receptor pathway inhibitors are more accurately predicted with NetBiTE. We hypothesize that the target genes for these drugs are more informative biomarkers of drug sensitivity than the targets of other drug categories. To further verify this hypothesis, we performed two control experiments: In one experiment, we compared NetBiTE and RF for cancer drugs that target a pathway that is not a membrane receptor pathway. We chose PI3K/MTOR signaling pathway inhibitors (PMSPIs), as PI3K/MTOR signaling pathway is a non-membrane tyrosine kinase pathway with a central role in cancer development (53). Moreover, PMSPIs are well represented in the GDSC database with 21 screened compounds. The results for this control experiment are shown in figure 5A and B. As shown in figure 5A, NetBiTE performs poorly across all $n_{tree}$ when compared to RF. Individual accuracy gains ($Δρ = ρ_{NB} – ρ_{RF}$) shown in figure 5B, are predominantly negative with only three of the 21 drugs showing a slight positive improvement ($Δρ$ < 0.05) in accuracy with NetBiTE. The observed results suggest that for PMSPI drugs, biasing the NetBiTE model towards the reported drug targets did not result in any accuracy gain, which in turn suggests that the expression of drug target genes for this category of drugs is not an informative biomarker for drug sensitivity.

In a second control experiment we selected two RSPI drugs: one that shows significant improvement in IC50 prediction accuracy with NetBiTE (Linifanib) and one that shows significantly worsened prediction accuracy with NetBiTE (PD173074). We then replaced the target genes for these two drugs with an equal number of randomly selected genes and applied NetBiTE to these new "dummy" drug targets. The results for this experiment are shown in figure 5C and D. Figure 5C shows that the use of dummy drug targets results in a reduced model accuracy that falls even below the RF accuracy for Linifanib. The results suggest that the randomly selected drug targets were uninformative, resulting in a poor predictive performance less than half of the prediction accuracy of NetBiTE with correct targets ($\rho = 0.3$ and $\rho = 0.7$ respectively). On the other hand, in figure 5D, the randomly assigned targets for PD173074 resulted in a prediction accuracy similar to that of RF. The results suggest that for PD173074, the drug targets reported in the literature were uninformative predictors for drug sensitivity and, as such, resulted in an inferior predictive performance when compared to randomly selected features.

Generally speaking, these results suggest that for those drugs on which NetBiTE performs poorly in predicting IC50 (i.e. non-responsive drugs), the expression of drug target genes reported in the literature is not an informative biomarker for IC50 drug sensitivity prediction. In the coming section, we will attempt to shed light on the reasons behind our observations in the context of cancer biology and mechanisms of drug sensitivity and resistance. Our preliminary explanation is that for drugs on which NetBiTE performs poorly, the mechanism of action may be through a complex biomolecular cascade that is independent of the expression of the target genes prior to drug administration. For these drugs as a result, there is no association or pattern between the target gene expression and IC50 drug sensitivity that can be identified using machine learning techniques such as NetBiTE.

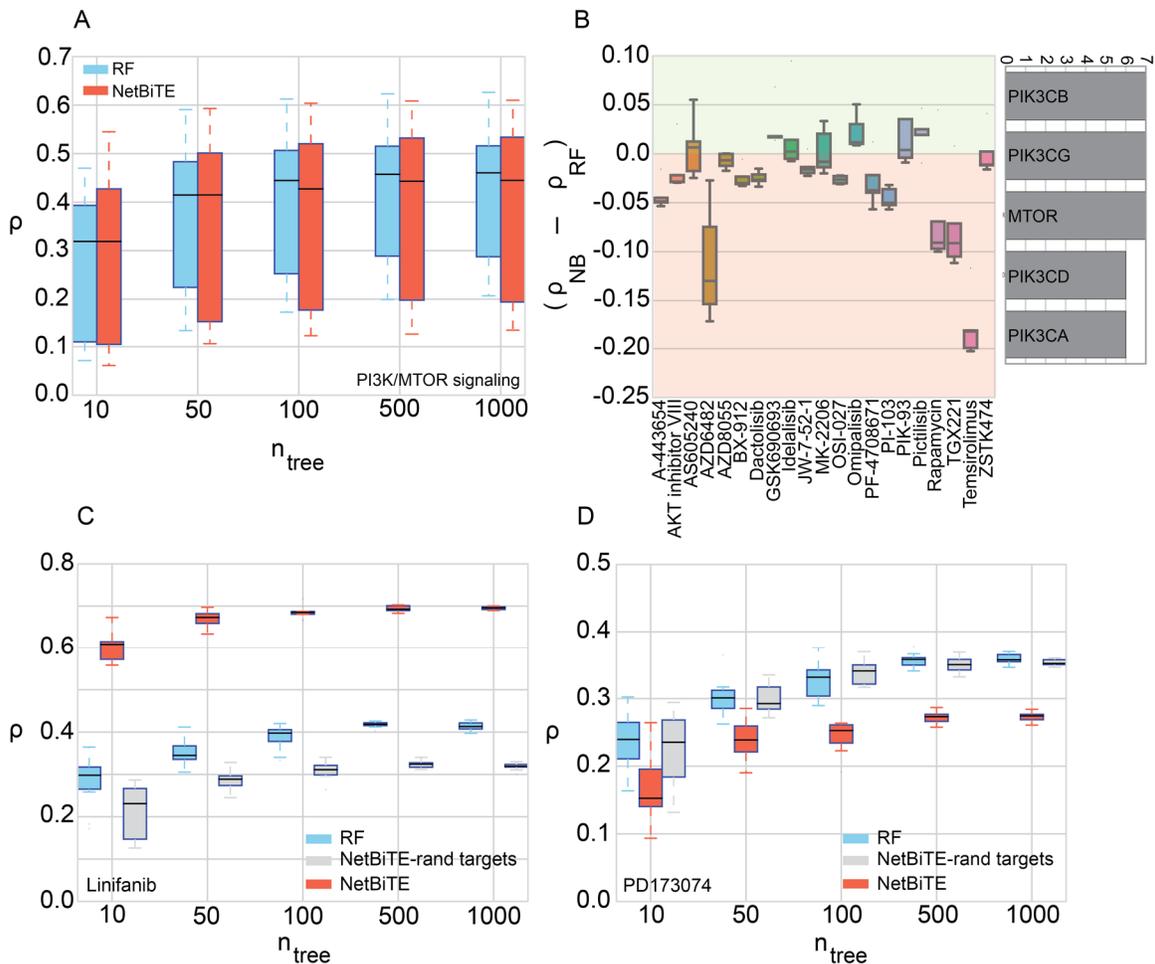

**Figure 5. Control experiments on PI3K/MTOR inhibitors and two control drugs with dummy targets confirm previous findings.** A) NetBiTE versus RF IC50 prediction results for 21 GDSC drugs that inhibit the PI3K/MTOR signaling pathway. As shown in the plot, NetBiTE performs either similarly or worse than RF across all drugs in this category and for all $n_{tree}$ studied. The lack of improvement in accuracy using NetBiTE suggests that the bias weights injected into the model were not informative and did enhance the model accuracy. B) Individual improvement in predictive performance for each drug in the category is presented along with a histogram showing the top five most frequently occurring target genes for the studied PMSPB drugs. Unlike the three MRPI drug categories, the vast majority of PI3K/MTOR signaling drugs demonstrate worse predictive performance with only three out of 21 drugs showing a slight improvement across all $n_{tree}$ with a mean improvement in accuracy below 0.05. C) Control experiment results for Linifanib where the true target genes were replaced with an equal number of random genes. As shown in the plot, the introduction of random targets results in a significant reduction in model performance compared to both RF (blue) and NetBiTE with true targets (red). D) Control experiment results for PD173074 where the true drug targets were replaced with random genes before NetBiTE was applied. As shown in the figure, introduction of random dummy targets resulted in a predictive performance similar to RF (blue) but superior to the predictive performance of NetBiTE with true targets (red).

Conversely, for drugs on which NetBiTE achieves high accuracy gains over RF in predicting IC50 (i.e. responsive drugs), the expression of target genes prior to drug administration appears to be a strong biomarker for IC50 drug sensitivity following drug administration. The results in figure 5 further confirm that the observed accuracy improvements with NetBiTE for MRPI drugs are due to the selection of highly informative genes and are not a result of random events, as substitution of the known drug targets by random dummy targets resulted in the loss of NetBiTE accuracy gains.

## 4. Discussion

We observed, from the NetBiTE results, that the expression of membrane receptor pathway genes in pan-cancer cell lines is predictive of IC50 drug sensitivity for compounds that inhibit those pathways. Interestingly, in a recent study of drug sensitivity in colorectal cancer for 16 cancer compounds, Schuette et al.(54) demonstrated experimentally that the sensitivity to EGFR inhibitors may be predicted using molecular biomarkers while they were unable to establish such a link for mTOR inhibitors. For instance, they observed that patient-derived organoid models with ERBB2 amplification were sensitive to drugs targeting ERBB2 (AZD8931 and Afatinib) but less so to drugs targeting EGFR only (Gefitinib) (54). In addition, it has been shown that ERBB2 amplification in breast cancer is a biomarker for sensitivity to Lapatinib, an ERBB2 inhibitor (3,8,55). This observation is in agreement with our identification of ERBB2 expression as a biomarker for drug sensitivity prediction (see supplementary figure S3).  Furthermore, RTK signaling enrichment and FLT3 mutations are a known biomarker for sensitivity to FLT3 inhibitors(47,56). In our studies, FLT3 was a frequent target in drugs that achieved enhanced accuracy with NetBite (see figure S3). Quizartinib, a novel second-generation FLT3 inhibitor with enhanced FLT3 specificity has shown favorable clinical outcomes in treating acute myeloid leukemia (56,57). Interestingly, Quizartinib is a top responder to NetBiTE (see figure 4B) highlighting the agreement of our findings with the literature where available.

Our findings regarding the MRP gene expression may be explained by the central role these pathways play in cancer development, survival and promotion of drug resistance. Growth factor signals are transmitted across the cell membrane, via their specific membrane receptors, to cytoplasmic signaling effectors that control many critical functions of cancer cells (58). In addition, various targeted therapy and drug resistance studies have pointed out that cancer cells appear to rely heavily on one or a few oncogenes or a single oncogenic pathway for survival and proliferation (59).This phenomenon is widely recognized as oncogenic addiction in cancer (5,59,60). Membrane receptor proteins have been shown to act as proto-oncogenes in various cancer types(61–63). For instance, hyper activation of receptor tyrosine kinases (RTKs, two-third of known TKs)(64) has been shown to be implicated in cancer even in the absence of extracellular activating ligands via overexpression of RTKs (65), activating mutations (66) or autocrine stimulation(53,67).

In addition, in many cancer cells where membrane receptor pathways are not the key oncogene, they appear to play a central role in promoting drug resistance by compensating for the inhibited pathway (68,69). For instance, in prostate cancer the inhibition of either of the key oncogenic pathways PI3K and AR, results in a feedback upregulation of the other pathway through induction of EGFR family RTK signaling (5).

One may pose the question, why the sensitivity to other drug categories such as PI3K/MTOR inhibitor drugs are not well predicted through their reported targets. We speculate that other drug categories that are non-responsive to NetBiTE, trigger complex cascade effects (54) that cannot be captured using only the expression of drug target genes, rendering drug sensitivity biomarker identification challenging. For instance, for non-membrane tyrosine kinases (NRTKs) it's been suggested that their involvement in cancer may not only be a result of over expression but also due to mutations or translocations (67) . As such we speculate that a more intricate epigenetic mechanism may be involved in the inhibitory mechanism of other drugs that are irresponsive to NetBiTE.

## 5. Conclusions

We have presented a new method and approach for drug sensitivity prediction and biomarker identification using a Network-based Biased Tree Ensembles (NetBiTE) algorithm that integrates prior knowledge and genetic profiles of cancer cells in order to make predictions. Prior knowledge comprises the known drug targets and a network of protein-protein interaction (PPI) such as the one provided by the STRING database. Using a synthetic dataset, we demonstrated that by identifying the most informative features and biasing split variable selection towards these features, Biased Tree Ensembles (BiTE) can significantly outperform the random forests (RF) algorithm given the same tuning parameters ($n_{tree}$ and *mtry*). In addition, we demonstrated that BiTE may be exploited for identification of biomarkers for cancer drug sensitivity. Next, we introduced NetBiTE, which combines BiTE with the propagation of drug target weights over the STRING PPI network. We used NetBiTE to study a panel of 50 cancer compounds tested with 883 pan-cancer cell lines within the GDSC dataset. The propagation of bias weights over the STRING PPI network simulates the underlying biological interactions between the target proteins and other key effectors following the drug perturbation. We showed that NetBiTE outperforms RF for the majority of drugs that inhibit membrane receptor pathway proteins, suggesting that the expression of membrane receptor pathway genes prior to drug administration is a biomarker for drug sensitivity of these compounds. To verify this observation, we performed two control studies: 1) using our model, we studied drugs that do not act on membrane receptor proteins. In this case, we chose drugs that block PI3K/MTOR pathway because of the prominent role of the pathway in cancer biology and the abundance of compounds within the GDSC database that target this pathway. NetBiTE did not result in improved accuracy when applied to drugs in this category, suggesting that unlike the MRPI drugs, the expression of target genes is not providing NetBiTE with any critical information that would result in an improvement in the model accuracy. 2) We performed a second control study in which we replaced the targets for two RTK signaling pathway inhibitors (Linifanib and PD173074), one responsive and one non-responsive to NetBiTE, with randomly assigned targets. Random assignment of targets resulted in significant worsening of the model performance in the case of Linifanib (the responsive drug) and no improvement at all compared to RF in case of PD173074 (the non-responsive drug). The results further confirmed that the performance improvement with NetBiTE is not a random occurrence, but is indeed the result of injecting informative prior-knowledge into the tree ensembles algorithm. We envision that our devised NetBiTE (and BiTE) method can play a key role as a readily implementable tool for testing the relevance of prior knowledge and for identifying critically informative features in various types of data and in particular in the context of personalized medicine for cancer, where identification of correct biomarkers is critical for the prediction of treatment outcomes.

**Acknowledgment**

The authors would like to thank Drs. Costas Bekas and Maria Gabrani for their continuous support and useful discussions. The project leading to this publication has received funding from the European Union's Horizon 2020 research and innovation programme under grant agreement No 668858.

**List of abbreviations**

RF = Random Forests

BiTE = Biased Tree Ensembles

NetBiTE = Network-based Biased Tree Ensembles

CART = Classification and Regression Trees

MRP = Membrane Receptor Pathway

MRPI = Membrane Receptor Pathway Inhibitor

RSPI = RTK Signaling Pathway Inhibitor

ESPI = EGFR Signaling Pathway Inhibitor

ISPI = IGFR Signaling Pathway Inhibitor

PMSPI = PI3K/MTOR Signaling Pathway Inhibitor

RTK = Receptor Tyrosine Kinase

EGFR = Epidermal Growth Factor Receptor

IGFR = Insulin-like Growth Factor Receptor

MTOR = Mammalian Target of Rapamycin

PI3K = Phosphoinositide 3-kinase

# Network-based Biased Tree Ensembles (NetBiTE) for Drug Sensitivity Prediction and Drug Sensitivity Biomarker Identification in Cancer

Ali Oskooei[1], Matteo Manica[1,2], Roland Mathis[1] and Maria Rodriguez Martinez[1]

[1] IBM Research – Zurich, Säumerstrasse 4, 8803 Rüschlikon, Switzerland

[2] Institute für Molekulare Systembiologie, Auguste-Piccard-Hof 1 8093, Zürich, Switzerland

**S1 Characterization of tuning parameters**

In this study, we swept across a range of values for number of trees ($n_{tree}$), *mtry* and the target partition size (*TPS*) to investigate how sensitive our model is to variations in each of these parameters when applying regression tree ensemble algorithms to the Genomics of Drug Sensitivity in Cancer (GDSC) (1) dataset. The tuning parameter characterization study was performed for a panel of 50 cancer drugs tested against 883 cell lines.

We studied the model performance with the number of trees ranging from 10 to 5000 as shown in figure S1B. As presented in the plot, beyond $n_{tree}$ = 50 the model performance is approximately flat with a p-value of 0.75 given by a one-way ANOVA test. Beyond 500 trees there is no detectable variation in model performances and as a result we adopted $n_{tree}$ = 500 as the optimal number of trees in our tree ensemble models.

In figure S1C, we studied the effect of *mtry* on model performance. We varied *mtry* from 10 to 5233, one-third of the total number of features (genes), which is the recommended mtry for RF regression in various sources in the literature (2,3). As shown in the plot the model performance does not vary across different *mtry* values with a p-value of 0.4 given by a one-way ANOVA test. This observation suggests that given the GDSC dataset, the choice of *mtry* within the recommended range, does not improve the model accuracy significantly. On the other hand, the choice of a high *mtry* is computationally expensive, as thousands of features need to be considered by the model rendering the model computationally inefficient. As such, the lowest possible *mtry* that results in a maximal model accuracy must be selected.

The effect of various target partition sizes (*TPS*) ranging from 1 to 100 was studied. As shown in the plot the RF algorithm is not sensitive to *TPS* as indicated by a p-value of 0.98 given by a one-way ANOVA test. As a result, we adopted *TPS* = 1 throughout this work as it resulted in the highest mean accuracy compared to other *TPS* values studied, however only by a slight margin.

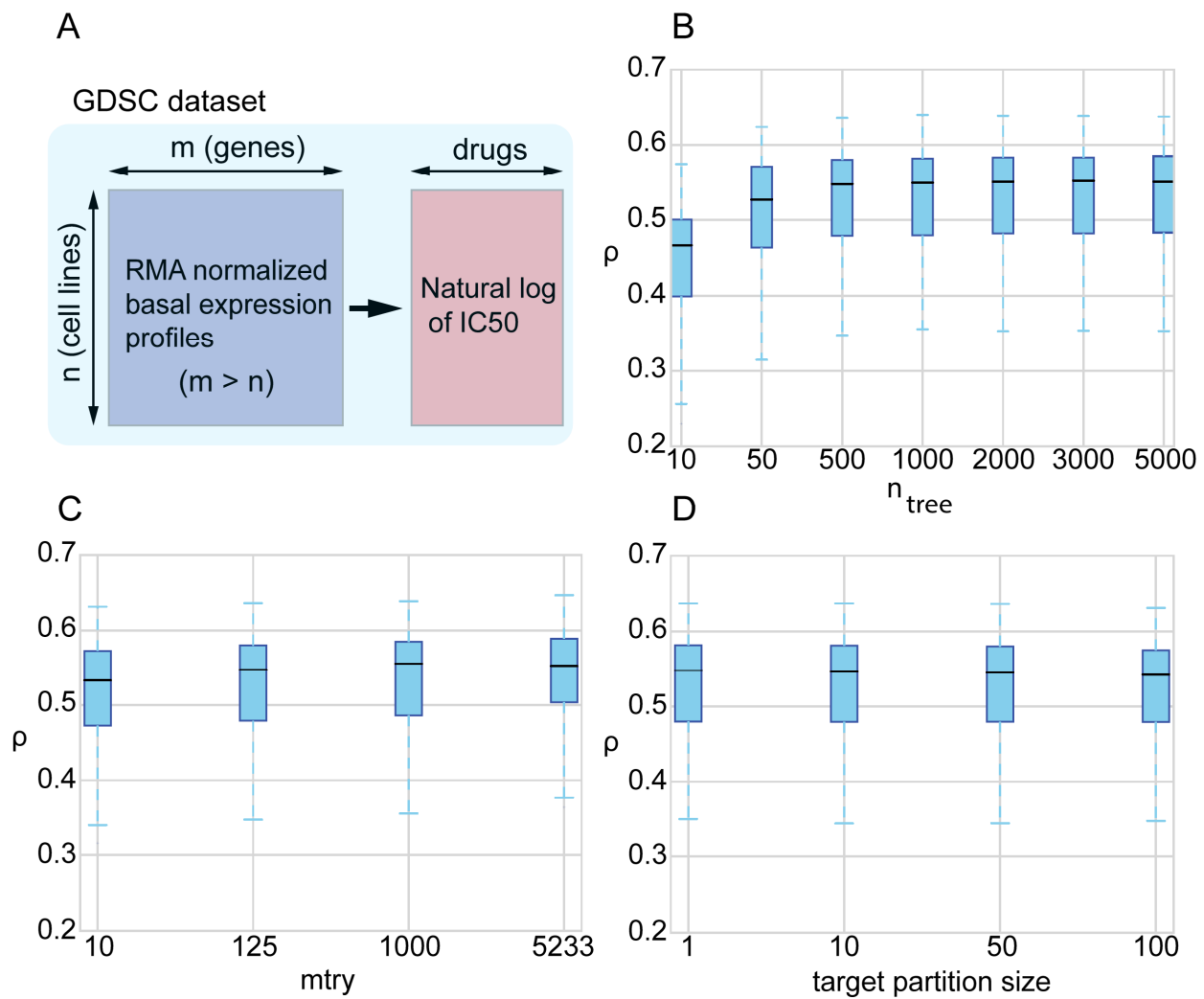

**Figure S1.** A) to train and test our models we used the drug sensitivity screening data available at Genomics of Drug Sensitivity in Cancer (GDSC) database (1). In our model, we incorporated RMA-normalized basal expression

profiles of *n* = 883 cancer cell lines tested with 50 compounds as well as the natural logarithm of IC50 for each compound tested with each cell line. B) using the GDSC dataset we trained a RF model with various number of trees ($n_{tree}$) while keeping *mtry* fixed at *mtry* = 125. As shown in the plot beyond 50 trees the model performance (Pearson correlation coefficients, $\rho$) plateaus with a p-value of 0.75 given by a one-way ANOVA test. C) Fixing the number of trees at $n_{tree}$ = 500, we investigated the effect of varying the number of features considered at each split, *mtry*, from *mtry* = 10 features to *mtry* = 5233 features. As shown in the plot the model performance was not sensitive to the choice of *mtry* as indicated by a one-way ANOVA test returning a p-value of 0.4. D) the effect of target partition size or the number of samples in each leaf node ($n_{tree}$ = 500 and *mtry* = 125), was investigated and was not found to impact the model significantly with a p-value of 0.98 given by a one-way ANOVA test. As a result, *TPS* = 1 was adopted throughout this work.

**S2 Investigation of propagation rate**

In figure S2, we investigated the effect of various α (i.e., the tuning parameter for the network diffusion depth of prior knowledge weights over the STRING network) for the drug Cabozantinib. As shown in the figure with a gradual increase in α from $\alpha$ = 0, the model accuracy increases until it peaks at $\alpha$ = 0.7. Any further increase in $\alpha$ past the optimal value of 0.7 results in a decrease in model accuracy. With this experiment, we confirmed that the optimal value of $\alpha$ = 0.7, as reported in literature for the STRING network, is valid. In addition, in this experiment we compared NetBiTE accuracy with the accuracy of RF and linear regression (LR) and observed that NetBiTE outperforms both RF and LR with a significant margin as shown in the figure S2.

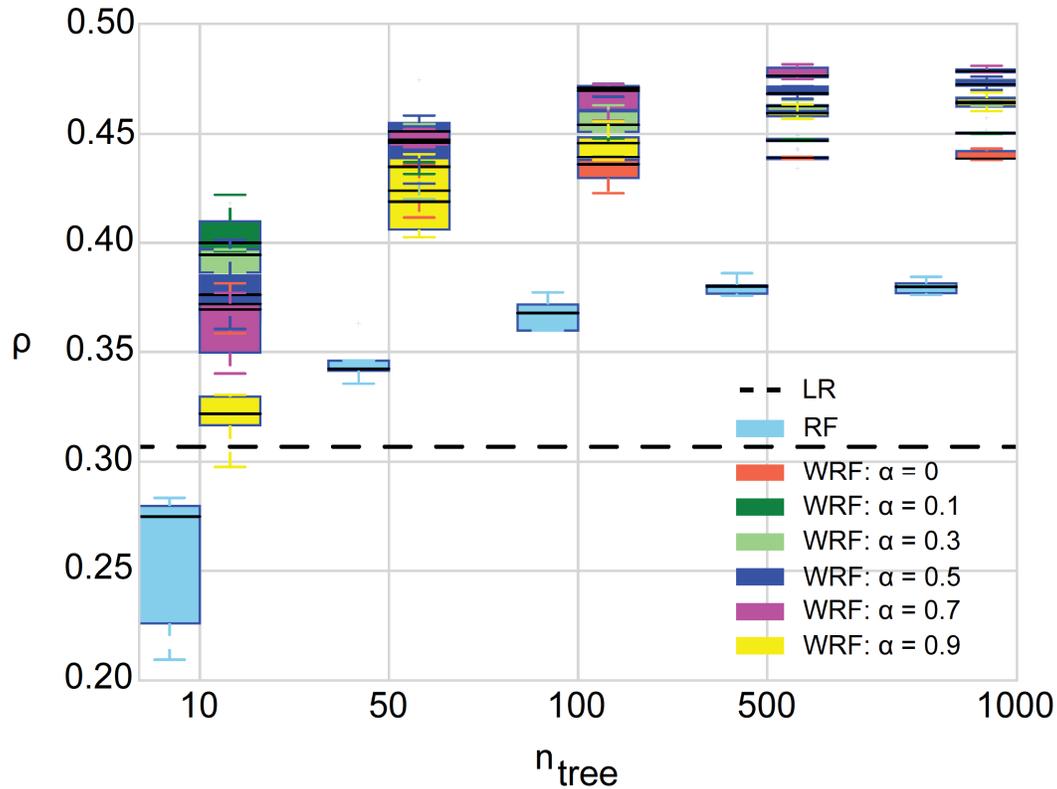

**Figure S2**. The study of variation in the network propagation tuning parameter $\alpha$ for Cabozantinib, one of the drugs that showed improvement with NetBiTE. As shown in the figure, the value for $\alpha$ was varied from 0 to 0.9 and the model performance was compared to standard RF and linear regression (LR). The model performance continuously increased with an increase in $\alpha$ and peaked at $\alpha = 0.7$. For $\alpha > 0.7$ NetBiTE performance dropped and at $\alpha = 1$ the weight vector was essentially uniform with equal weights for all genes resulting in a NetBiTE performance that is identical to standard RF. As such, we confirmed $\alpha = 0.7$ as the optimal value and chose this value throughout our studies.

**S3 Analysis of drug target genes**

In order to determine which membrane receptor genes are more likely to be responsible for the accuracy improvements observed using NetBiTE (see figure 4), we plotted the top three genes and their frequency of occurrence in responsive and non-responsive drugs within each category. For RTK signaling pathway inhibitors (RSPIs), there is a clear discrepancy between the most frequent genes in responsive versus non-responsive drugs, as shown in figure S3A and S3B. KIT, FLT3 and PDGFRB are the most frequent targets in responsive drugs while KDR, TGFBR1 and FGFR1 are most frequent targets in non-responsive RSPI drugs. For EGFR signaling pathway inhibitors (ESPIs), ERBB2 is a target exclusively in responsive drugs. For IGFR signaling pathway inhibitors, INSRR is a frequent target, exclusively in responsive drugs.

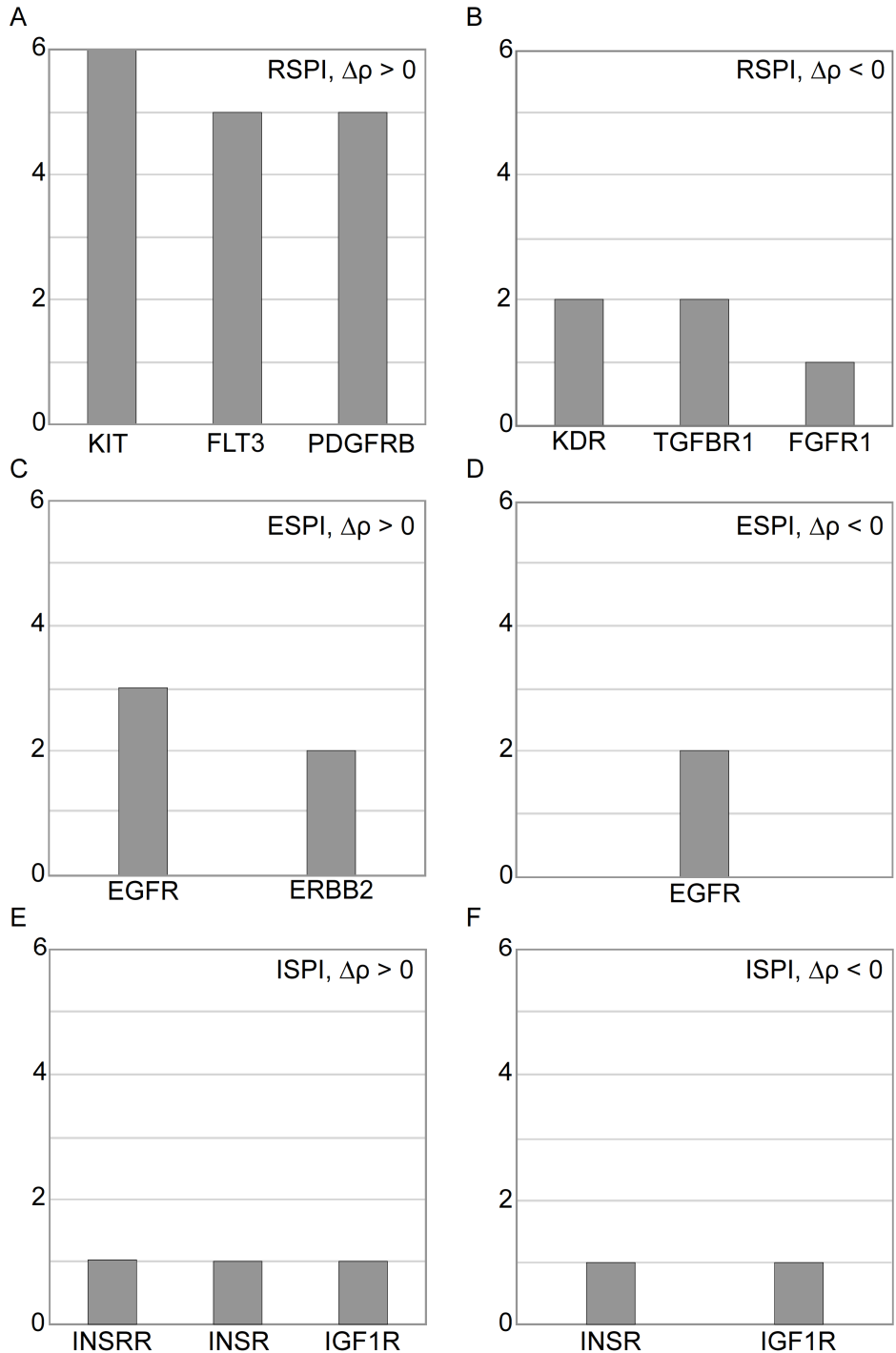

**Figure S3.** Top-three most frequent target genes in drugs that are responsive ($\Delta\rho > 0$) and non-responsive ($\Delta\rho < 0$) to NetBiTE for the three categories of membrane receptor inhibitor drugs studied. The values on the y-axis are the number of drugs that had the given gene as a target. A) Most frequent genes among responsive RTK signaling pathway inhibitors. B) Most frequent genes among non-responsive RTK signaling pathway inhibitors. C) Most frequent genes among responsive EGFR signaling pathway inhibitors. D) Most frequent genes among non-responsive EGFR signaling pathway inhibitors. E) Most frequent genes among responsive IGFR signaling pathway inhibitors. F) Most frequent genes among non-responsive IGFR signaling pathway inhibitors.